\documentclass[a4paper]{article}

\usepackage[utf8]{inputenc}

\usepackage{multicol}
\usepackage{amsthm}
\usepackage{bm}
\usepackage{physics}

\usepackage[dvipdfmx]{graphicx}

\usepackage{bmpsize}
\usepackage{amssymb}
\usepackage{amsmath}
\usepackage{color}
\usepackage{amsthm}
\usepackage{comment}

\newcommand{\BA}{\begin{eqnarray}}
\newcommand{\EA}{\end{eqnarray}}

\definecolor{dgreen}{rgb}{0.0, 0.5, 0.0}

\begin{document}

\fontsize{14pt}{16.5pt}\selectfont

\begin{center}
\bf{
Types and stability of fixed points for positivity-preserving discretized dynamical systems in two dimensions
}
\end{center}
\fontsize{12pt}{11pt}\selectfont
\begin{center}
Shousuke Ohmori$^{1, *)}$ and Yoshihiro Yamazaki$^2$\\ 
\end{center}

\noindent
$^1$\it{National Institute of Technology, Gunma College, Maebashi-shi, Gunma 371-8530, Japan}\\
$^2$\it{Department of Physics, Waseda University, Shinjuku, Tokyo 169-8555, Japan}\\

\noindent
*corresponding author: 42261timemachine@ruri.waseda.jp\\
~~\\
\rm
\fontsize{11pt}{14pt}\selectfont\noindent

\baselineskip 30pt

\noindent
{\bf Abstract}\\
%
%
Relationship for dynamical properties in the vicinity of fixed points 
between two-dimensional continuous and its positivity-preserving discretized dynamical systems is studied.
Based on linear stability analysis, 
we reveal the conditions under which the dynamical structures of the original continuous dynamical systems are retained in their discretized dynamical systems, 
and the types of fixed points are identified if they change due to discretization.
We also discuss stability of the fixed points in the discrete dynamical systems.
The obtained general results are applied to Sel'kov model and Lengyel-Epstein model.

\bigskip

\bigskip




%
%
\section{Introduction}

Many studies have been conducted on the dynamical properties of ultradiscrete equations derived from continuous differential equations for non-integrable systems, such as reaction-diffusion systems\cite{Murata2013,Matsuya2015,Ohmori2016}, an inflammatory response system\cite{Carstea2006,Willox2007} and a biological negative-feedback system\cite{Gibo2015}.
%
For these derivations, it is necessary to adopt a difference method that preserves the positivity of the continuous equations in order to apply the ultradiscrete limit\cite{{Tokihiro1996}}.
As one of the positivity-preserving difference methods, 
the tropical discretization method\cite{{Murata2013}} is often applied.
Then, understanding correspondence of dynamical properties among original continuous differential equations, their tropically discretized equations and their ultradiscretized equations is important, 
and whether the derived discrete equations can retain the dynamical properties of the original models is a significant problem.

With the above problem in mind, we have recently studied dynamical properties of the local bifurcations (transcritical, saddle-node, pitchfork) in one-dimensional dynamical systems\cite{Ohmori2020,Ohmori2022b}.
In these previous studies, we argued stabilities of fixed points in continuous differential equations and their tropically discretized ones.
%
We successfully identified conditions under which the derived discrete dynamical systems can retain the bifurcations of the original continuous ones via tropical discretization.

In this letter, we focus on types and stability of fixed points in two-dimensional dynamical systems.
So far, we have numerically investigated Sel'kov model as a specific example\cite{Yamazaki2021,Ohmori2021,Ohmori2022a}.
However, the previous studies have not been treated 
as generally and analytically as in the above one-dimensional case.
Moreover in two-dimensional dynamical systems, there is an essential difference from the one-dimensional case: existence of Hopf (or Neimark-Sacker) bifurcation.
Therefore, in addition to the general treatment of one-dimensional cases, that of two-dimensional cases is meaningful and important. 
%
%
%
\section{General results}

Now we consider the following continuous two-dimensional differential equation with positive variables ${\mathbf {x}}=(x_1,x_2)$
%
%
\begin{eqnarray}
	\frac{d\mathbf{x}}{dt}& = \mathbf{F}(\mathbf{x})=\mathbf{f}(\mathbf{x})-\mathbf{g}(\mathbf{x}),
	\label{eqn:1-1} 
\end{eqnarray} 
where $x_1,x_2>0$ and 
we assume that $\mathbf{F}(\mathbf{x})$ can be divided into two positive smooth functions $\mathbf{f}=(f_1,f_2)$ and $\mathbf{g}=(g_1,g_2)$.
By the tropical discretization\cite{Murata2013}, 
we obtain the following discrete dynamical system 
from eq.(\ref{eqn:1-1}):  
%
\begin{eqnarray}
    \mathbf{x} _{n+1}=\mathbf{x} _n \frac{\mathbf{x} _n+\tau \mathbf{f}(\mathbf{x} _n)}{\mathbf{x} _n+\tau \mathbf{g}(\mathbf{x} _n)},
    \label{eqn:1-2}
\end{eqnarray}
where $\tau (> 0) $ and $n$ show the discretized time interval and 
the number of iteration steps with non-negative integer;
$\mathbf{x}_n = \mathbf{x}(n \tau)$, respectively.
%
%
%
%
It is found that if $\bar {\mathbf{x}}=(\bar x_1,\bar x_2)$ is a fixed point of eq.(\ref{eqn:1-1}), 
then $\bar {\mathbf{x}}$ also becomes a fixed point of eq.(\ref{eqn:1-2}).
Hereafter the fixed points $\bar {\mathbf{x}}$ of 
eqs.(\ref{eqn:1-1}) and (\ref{eqn:1-2}) are denoted by 
$\bar {\mathbf{x}}^{(c)}$ and $\bar {\mathbf{x}}^{(d)}$, respectively,  
when their difference is needed to be made clear.
We set the Jacobian $\mathbf{J(\bar {\mathbf{x}})}=
\begin{pmatrix}
D_{11} & D_{12} \\
D_{21} & D_{22} \\
\end{pmatrix}$ 
at $\bar {\mathbf{x}}^{(c)}$, where 
$D_{ij}
=\displaystyle\frac{\partial{(f_i-g_i)}}{\partial{x_j}}(\bar {\mathbf{x}})$ ($i,j=1,2$).
The trace $T$ and the determinant $\Delta$ of $\mathbf{J(\bar {\mathbf{x}})}$ is obtained as $T = D_{11}+D_{22}$ and $\Delta=D_{11}D_{22}-D_{12}D_{21}$.
The Jacobian of eq.(\ref{eqn:1-2}) is also given by 
$\mathbf{J_\tau (\bar {\mathbf{x}})}=
\begin{pmatrix}
1 + Z_{\tau 1}D_{11} & Z_{\tau 1}D_{12} \\
Z_{\tau 2}D_{21} & 1 + Z_{\tau 2}D_{22} \\
\end{pmatrix}$, 
where 
$Z_{\tau i}\equiv \displaystyle\frac{\tau \bar x_i}{\bar x_i + \tau \bar f_i}~(>0)$ 
and $\bar f_i\equiv f_i(\bar {\mathbf{x}})$ ($i=1,2$).
Note that the trace $T '$ and the determinant $\Delta '$ of $\mathbf{J_\tau (\bar {\mathbf{x}})}$ are given as 
\begin{eqnarray}
    T'
     & = & 2+\displaystyle\frac{\tau \bar x_1 \bar x_2T+\tau^2(\bar x_1\bar f_2 D_{11}+\bar x_2\bar f_1 D_{22})}{(\bar x_1 + \tau \bar f_1)(\bar x_2 + \tau \bar f_2)}, 
   \nonumber \\ 
%
    \Delta '
    &=&T'-1+Z_{\tau 1}Z_{\tau 2}\Delta.
    \label{eqn:2-1-1}
\end{eqnarray}
%
%
\subsection{Type of the fixed point}

\begin{figure}[h!]
    \begin{center}
	\includegraphics[width=10cm]{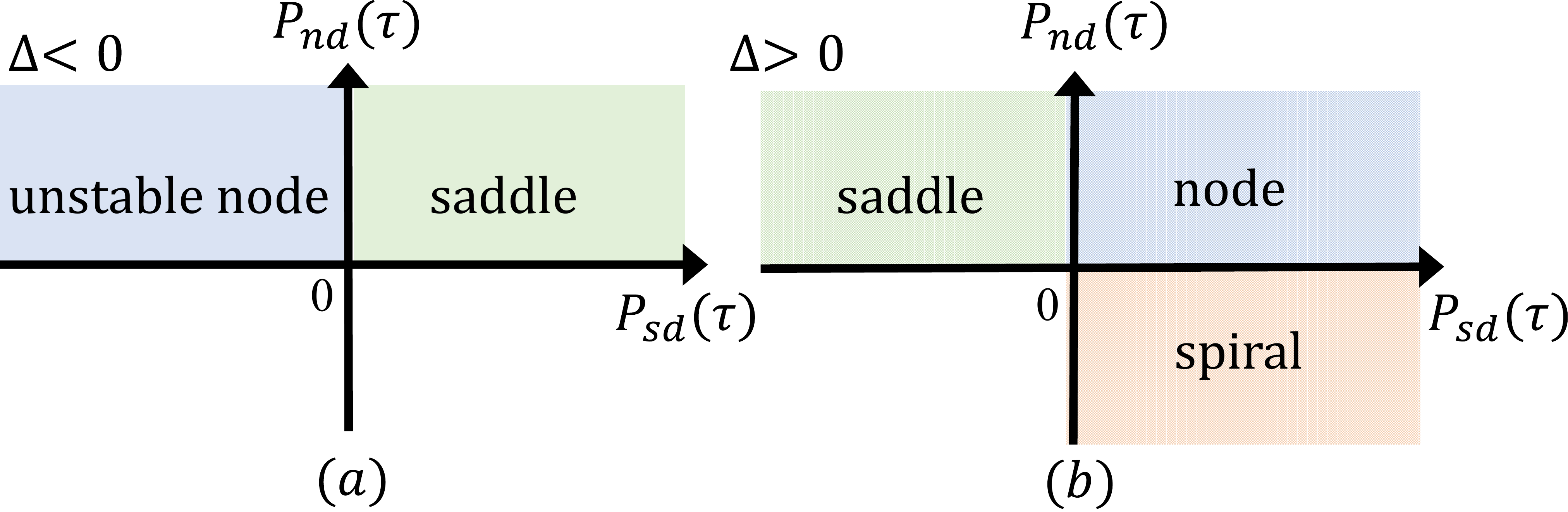}
    \caption{\label{Fig.Classification} Classifications of the type of the fixed point in eq.(\ref{eqn:1-2}), $\bar {\mathbf{x}}^{(d)}$. 
    (a) $\Delta <0$. (b) $\Delta >0$. 
    %
    %
   }
   \end{center}
\end{figure}

\begin{figure}[h!]
    \begin{center}
    \includegraphics[width=8cm]{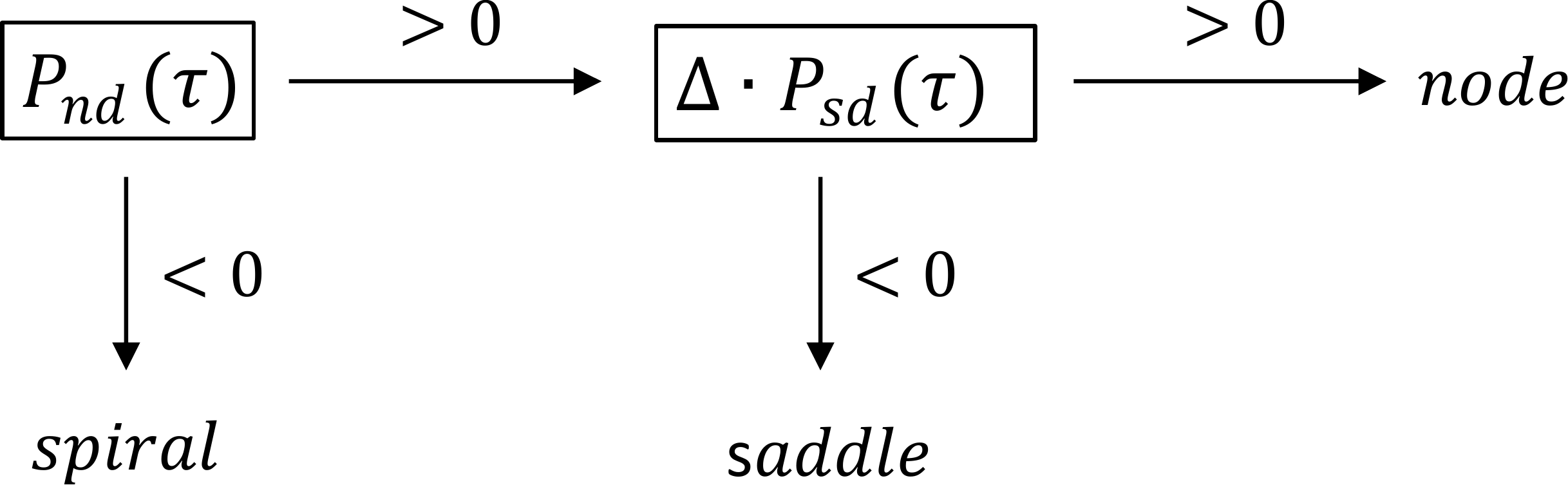}
    \caption{\label{Fig_Flowchart} Flowchart for obtaining the type of $\bar {\mathbf{x}}^{(d)}$.}
    \end{center}
\end{figure}

Focusing on the sign of $\Delta$, 
we determine relationship of the three types of fixed points 
(saddle, node, spiral) between in two dimensional continuous and discrete dynamical systems\cite{Strogatz2014,Galor}.
(i) When $\Delta <0$, $\bar {\mathbf{x}}^{(c)}$ becomes a saddle.
From eq.(\ref{eqn:2-1-1}), $\Delta '< T '-1$ holds.
Then, $\bar {\mathbf{x}}^{(d)}$ becomes either a saddle or an unstable node.
In fact, $\bar {\mathbf{x}}^{(d)}$ becomes saddle when $\Delta ' > -T '-1$, otherwise unstable node.
Expressing $T '$ and $\Delta '$ in terms of $\tau$,
we have the quadratic inequality corresponding to $\Delta ' > -T ' -1$, 
\begin{equation}
  P_{sd}(\tau) \equiv A_{sd} \tau^2 + B_{sd} \tau + C_{sd} > 0,
  \label{eqn:saddle_ineq}
\end{equation}
where 
\begin{eqnarray}
  A_{sd} & = & 2(\bar x_1 \bar f_2 D_{11}+\bar x_2 \bar f_1 D_{22})+\bar x_1 \bar x_2 \Delta +4\bar f_1 \bar f_2, \nonumber \\
  B_{sd} & = & 2 \bar x_1\bar x_2 T +4(\bar x_1\bar f_2+\bar x_2\bar f_1), \\ 
  C_{sd} & = & 4\bar x_1\bar x_2 \;\;\; (> 0). \nonumber
  \label{eqn:saddle_ineq_1} 
\end{eqnarray}
By using $P_{sd}(\tau)$, the above statement can be rewritten as follows; $\bar {\mathbf{x}}^{(d)}$ becomes saddle 
for $P_{sd}(\tau) > 0$ and unstable node for $P_{sd}(\tau) < 0$ when $\Delta <0$. 
(ii) In the case of $\Delta >0$,  
the fixed point $\bar {\mathbf{x}}^{(c)}$ becomes node when $4\Delta < T^2$ and spiral when $4\Delta > T^2$.
From $\Delta >0$, we obtain $\Delta '>T '-1$ 
based on eq.(\ref{eqn:2-1-1}).
In this case, the fixed point $\bar {\mathbf{x}}^{(d)}$ can be classified as follows.
(ii-a) When $ P_{sd}(\tau) > 0$ and $4\Delta'<{T'} ^2$, 
$\bar {\mathbf{x}}^{(d)}$ becomes node.
Note that the inequality $4\Delta'<{T'} ^2$ is transformed into the following quadratic inequality with respect to $\tau$ by eq.(\ref{eqn:2-1-1}): 
\begin{equation}
  P_{nd}(\tau) \equiv A_{nd} \tau^2  + B_{nd} \tau + C_{nd} > 0,
  \label{eqn:spiral_ineq}  
\end{equation}
where 
\begin{eqnarray}
 \label{eqn:spiral_ineq_1} 
 A_{nd} & = & (\bar x_1\bar f_2D_{11}+\bar x_2\bar f_1D_{22})^2
  -4\bar x_1\bar x_2\bar f_1\bar f_2\Delta, \nonumber \\
  B_{nd} & = &  2\bar x_1\bar x_2T (\bar x_1\bar f_2D_{11}+\bar x_2\bar f_1D_{22})\nonumber \\
  & & -4\bar x_1\bar x_2(\bar x_1\bar f_2+\bar x_2\bar f_1)\Delta, \\ 
  C_{nd} & = & (\bar x_1\bar x_2)^2(T^2 - 4\Delta). \nonumber
\end{eqnarray}
%
(ii-b) When $ P_{sd}(\tau) > 0$ and $4\Delta' > {T'} ^2$, 
or $P_{nd}(\tau) < 0$, 
$\bar {\mathbf{x}}^{(d)}$ becomes spiral. 
(ii-c) When $ P_{sd}(\tau) < 0$, $P_{nd}(\tau) > 0$ automatically holds 
and $\bar {\mathbf{x}}^{(d)}$ becomes saddle.
In summary, the type of the fixed point $\bar {\mathbf{x}}^{(d)}$ can be determined 
depending on the sign of $\Delta$, $P_{sd}(\tau)$, and $P_{nd}(\tau)$ as shown in Fig.\ref{Fig.Classification}.
Note that $P_{nd}(\tau)>0$ automatically holds 
for any $\tau$ when $\Delta < 0$.
In a simpler way, the type of $\bar {\mathbf{x}}^{(d)}$ can be determined according to the flowchart shown in Fig.\ref{Fig_Flowchart}.

\subsection{Stability}

Here we consider the stability of $\bar {\mathbf{x}}^{(d)}$.
Assuming $\Delta '>\pm T'-1$, the fixed point $\bar {\mathbf{x}}^{(d)}$ becomes stable when the inequality $T'<1$ is satisfied.
From eq. (\ref{eqn:2-1-1}), $T'<1$ is equivalent to the inequality,
\begin{equation}
  \alpha (\bar {\mathbf{x}}) \tau + \beta(\bar {\mathbf{x}}) < 0, 
  \label{eqn:stable_ineq}
\end{equation}
where 
\begin{eqnarray}
    \alpha (\bar {\mathbf{x}})  =  \Delta\bar x_1 \bar x_2+(\bar x_1 \bar f_2 D_{11}+\bar x_2 \bar f_1 D_{22}), 
    \beta(\bar {\mathbf{x}})  =  T\bar x_1 \bar x_2.
    \label{eqn:stable_ineq_ab}
\end{eqnarray}
%
%
Solving this inequality, we find the following cases for stability of $\bar {\mathbf{x}}^{(d)}$.
%
[St-1] When $\alpha (\bar {\mathbf{x}}) > 0$, 
we obtain $\displaystyle \tau < -\frac{\beta (\bar {\mathbf{x}})}{\alpha (\bar {\mathbf{x}})} (\equiv \gamma(\bar {\mathbf{x}}))$ in which $\bar {\mathbf{x}}^{(d)}$ is stable. 
Note that when $\beta (\bar {\mathbf{x}})$ is positive, 
$\bar {\mathbf{x}}^{(d)}$ is unstable for any $\tau >0$.
%
[St-2] When $\alpha (\bar {\mathbf{x}}) < 0$, 
$\displaystyle \tau > -\frac{\beta (\bar {\mathbf{x}})}{\alpha (\bar {\mathbf{x}})}$ is obtained for the condition of $\tau$ under which $\bar {\mathbf{x}}^{(d)}$ is stable. 
Note that when $\beta (\bar {\mathbf{x}})$ is negative, 
$\bar {\mathbf{x}}^{(d)}$ is stable for any $\tau >0$.
\section{Applications}

Here we demonstrate application of the above general results 
to the following two examples: 
(1) Sel'kov model and (2) Lengyel-Epstein model.\\
%
%
(1) \textit{Sel'kov model}\cite{Strogatz2014, Selkov1968}: 
\begin{eqnarray}
    \frac{dx}{dt}  =  -x+ay+x^2y,~
    \frac{dy}{dt}  =  b-ay-x^2y,
    \label{eqn:0-1-1-cont}
\end{eqnarray}
where $a$ and $b$ are positive bifurcation parameters. 
From eq.(\ref{eqn:0-1-1-cont}), it is found that the fixed point $\displaystyle \bar {\mathbf{x}}^{(c)}=\left(b,\frac{b}{a+b^2}\right)$ is spiral.
Setting 
\begin{eqnarray}
    f_1(x,y) & = & ay + x^2 y, ~ 
    g_1(x,y) = x, ~\nonumber \\
    f_2(x,y) & = & b, ~
    g_2(x,y) = ay + x^2 y,
    \label{eqn:4-1-fg}  
\end{eqnarray}
we obtain
\begin{eqnarray}
    T  =  -\frac{b^4+(2a-1)b^2+a+a^2}{a+b^2},~ \Delta=a+b^2, \nonumber\\
    D_{11}  =  -1+\frac{2b^2}{a+b^2},~ D_{22}=-(a+b^2),~ 
    \bar f_1  =  \bar f_2=b.
    \label{eqn:2-2-1}
\end{eqnarray}
Then $4 \Delta > T^{2} \; (>0)$ holds 
and the following discretized equation is obtained 
via tropical discretization\cite{Murata2013,Ohmori2021},
\begin{eqnarray}
    x_{n+1}   =  \frac{x_n+\tau (ay_n+x^2_ny_n)}{1+\tau},
    y_{n+1}   =  \frac{y_n+\tau b}{1+\tau (a+x^2_n)}.
    \label{eqn:0-1-1}
\end{eqnarray}
From eq.(\ref{eqn:spiral_ineq_1}), we find  
$A_{nd}=-\displaystyle\frac{4(2ab^6+b^8)}{(a+b^2)^2} < 0$, 
$B_{nd}=-\displaystyle\frac{4 b^6 (a^2 + b^2 + b^4 + a (3 + 2 b^2))}{(a + b^2)^3}<0$, 
$C_{nd} < 0$.
Therefore, $P_{nd}(\tau) < 0$ for all $\tau > 0$ and  
the fixed point of eq.(\ref{eqn:0-1-1}) is also spiral 
for any $\tau >0$.

\begin{figure}[h!]
    \begin{center}
    \includegraphics[width=10cm]{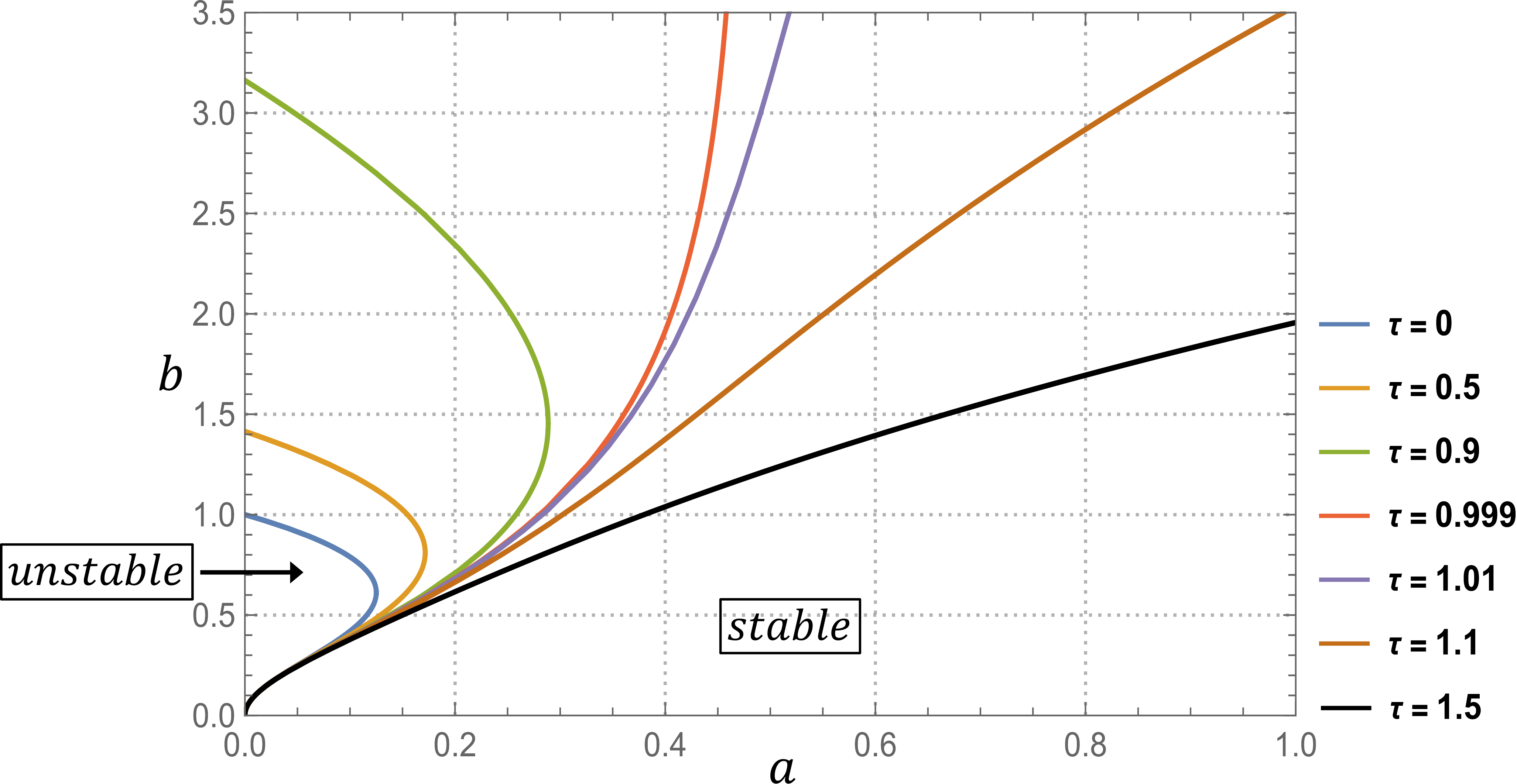}
    \caption{\label{Fig.bif_ab_curve} $\tau $-dependence of the bifurcation curves, eq.(\ref{eqn:2-4-2}). }
    \end{center}
\end{figure}

\begin{figure}[h!]
    \begin{center}
    \includegraphics[width=10cm]{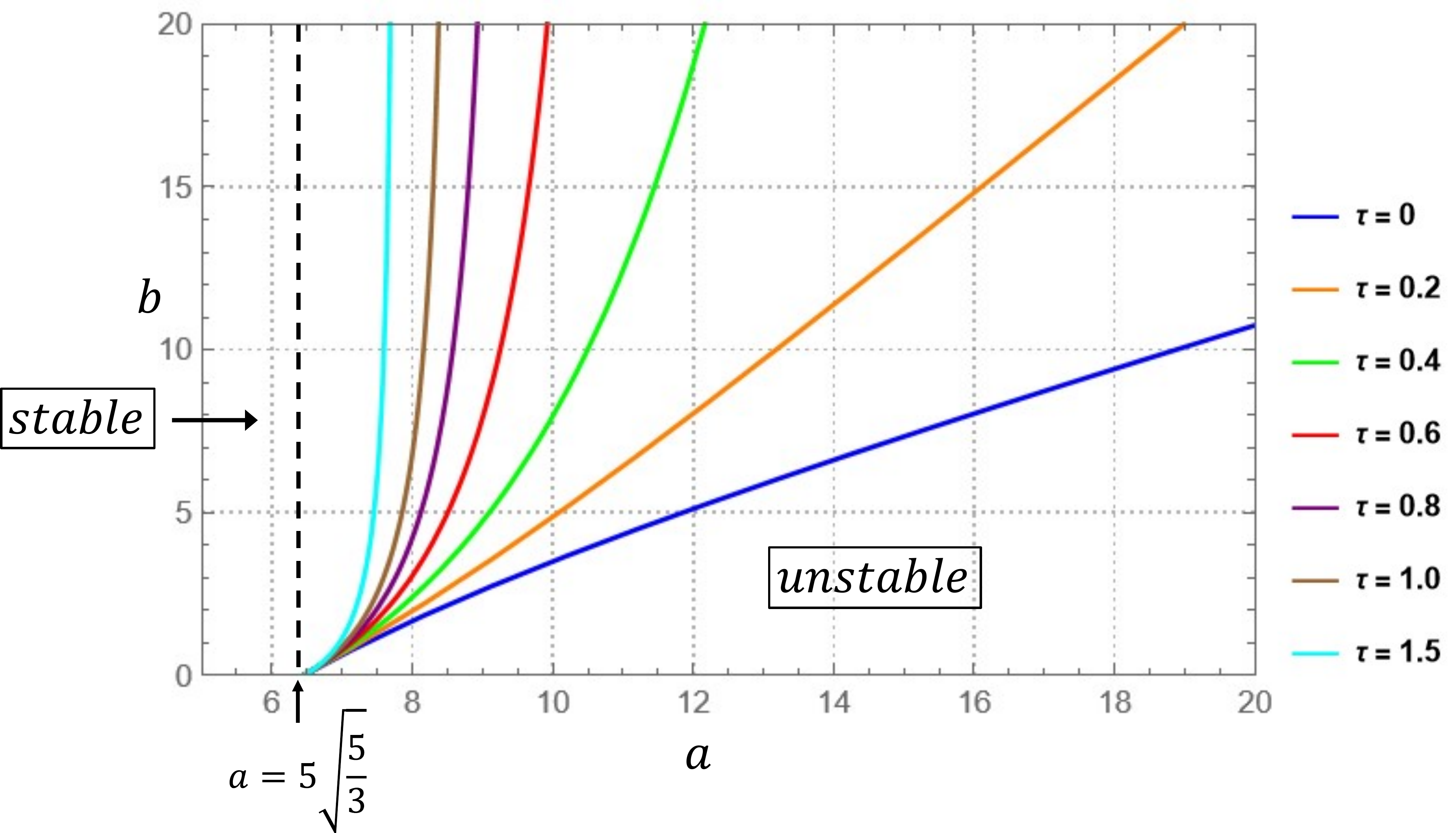}
    \caption{\label{Fig.LE_bif_ab_curve} $\tau$-dependence of the bifurcation curves, 
    eq.(\ref{eqn:4-5}). }
    \end{center}
\end{figure}

Next we focus on stability of the spiral fixed point $\displaystyle \bar {\mathbf{x}}^{(d)} = \left( b,\frac{b}{a+b^2} \right)$ of eq.(\ref{eqn:0-1-1}).
In this case, $\Delta '>\pm T'-1$ is satisfied.
From eqs.(\ref{eqn:stable_ineq_ab}) and (\ref{eqn:2-2-1}), we obtain
$\alpha (\bar {\mathbf{x}}) = \displaystyle\frac{b^2(b^2-a)}{a+b^2}$
and
$\displaystyle \beta (\bar {\mathbf{x}}) = 
-\frac{b^2\{(b^2 + a)^2 + a - b^2\}}{(a+b^2)^2}$.
If $\alpha (\bar {\mathbf{x}})<0$, or $b < \sqrt{a}$, 
then $\beta (\bar {\mathbf{x}})<0$.
Therefore from [St-2], $\bar {\mathbf{x}}^{(d)}$ is stable
in eq.(\ref{eqn:0-1-1}) for any $\tau>0$. 
If $\alpha (\bar {\mathbf{x}}) > 0$, or $b > \sqrt{a}$, 
$\bar {\mathbf{x}}^{(d)}$ becomes stable from [St-1] for $0 < \tau < \gamma$ 
and unstable for $\tau > \gamma$.
Here $\gamma $ is given 
as $\displaystyle \gamma =\frac{b^4+(2a-1)b^2+a+a^2}{b^4-a^2}$.
%
%
%
%
%
%
Therefore $\gamma = \tau$ gives the bifurcation surface 
for Neimark-Sacker (Hopf) bifurcation of $\bar {\mathbf{x}}^{(d)}$.
Actually from $\gamma = \tau$, we obtain 
%
%
\begin{equation}
    b^2 = \frac{1-2a \pm\sqrt{1-8a+4a \tau +4a^2 \tau ^2}}{2(1-\tau)}.
    \label{eqn:2-4-2}   
\end{equation}
Figure \ref{Fig.bif_ab_curve} shows the projection of the bifurcation surface given by eq.(\ref{eqn:2-4-2}) to the $ab$-plane.
From this figure, the following features are confirmed.
(i) When $\tau \to 0$, eq.(\ref{eqn:2-4-2}) provides  $\displaystyle b^2=\frac{1}{2}(1-2a\pm \sqrt{1-8a})$, which coincides with the bifurcation curve of eq.(\ref{eqn:0-1-1-cont}).
(ii) When $\tau \to \infty$, 
eq.(\ref{eqn:2-4-2}) becomes $b = \sqrt{a}$.
(iii) For any $\tau$, we can find the bifurcation curve 
in the $ab$-plane.
This feature suggests that for any $\tau>0$ the Neimark-Sacker bifurcation occurs.
Note that the bifurcation surface, eq.(\ref{eqn:2-4-2}),  analytically reproduces that obtained by numerical calculation in the previous study \cite{Ohmori2021}.
%
%
%
%
%
%
%
\\
(2) \textit{Lengyel-Epstein model}\cite{Strogatz2014,Lengyel}: 
\begin{eqnarray}
    \frac{dx}{dt}  =  a-x-\frac{4xy}{1+x^2},~
    \frac{dy}{dt}  =  bx \left( 1-\frac{y}{1+x^2} \right),
    \label{eqn:4-1}
\end{eqnarray}
where $a$ and $b$ are positive bifurcation parameters.
In eq.(\ref{eqn:4-1}), the Hopf bifurcation occurs 
when $\displaystyle b=\frac{3}{5}a-\frac{25}{a}$.
The spiral fixed point of eq. (\ref{eqn:4-1}) is 
$\displaystyle \bar {\mathbf{x}}^{(c)}=\left( \frac{a}{5},1+\left( \frac{a}{5} \right)^2 \right) (=(\bar x, 1 + \bar x^2))$.
When $\displaystyle b < \frac{3}{5}a-\frac{25}{a}$, 
the fixed point $\bar {\mathbf{x}}^{(c)}$ becomes unstable 
and the limit cycle emerges around $\bar {\mathbf{x}}^{(c)}$.
For application of eq.(\ref{eqn:1-1}) to eq.(\ref{eqn:4-1}), we divide the right hand sides of eq.(\ref{eqn:4-1}) into the positive smooth functions $f_i$ and $g_i$ ($i=1,2$): 
\begin{eqnarray}
    f_1(x,y) & = & a,~ g_1(x,y)  =  \frac{4xy}{1+x^2}+x,\nonumber \\
    f_2(x,y) & = & bx,~ g_2(x,y)  =  \frac{bxy}{1+x^2}.
    \label{eqn:4-2}   
\end{eqnarray}
Then the tropically discretized equation 
of the Lengyel-Epstein model is obtained as
\begin{eqnarray}
    x_{n+1}  =  \frac{x_n+\tau a}{1+\tau(1+\frac{4 y_n}{1+x_n^2})},~
    y_{n+1}  =  \frac{y_n+\tau b x_n}{1+\tau \frac{b x_n}{1+x_n^2}}.
    \label{eqn:4-3}
\end{eqnarray}
%
%
%
%
Here we focus on the type of $\bar {\mathbf{x}}^{(d)}$
in eq.(\ref{eqn:4-3}).
Since 
$\displaystyle 
T  =  \frac{3 \bar{x} ^2-b \bar x-5}{1+\bar{x}^2},~ \Delta=\frac{5 b \bar x}{1+\bar{x}^2} (>0) ,~\bar f_1=a, ~ \bar f_2=b \bar x,     D_{11}  =  \frac{3\bar x ^2-5}{1+\bar{x}^2}, ~ D_{22}=-\frac{b \bar x}{1+\bar{x}^2}$,
we obtain $4\Delta - T^{2} > 0$.
And we find 
$\displaystyle A_{nd}=-\frac{32b^2 \bar x ^6 (5+3 \bar x ^2)}{(1+\bar x ^2)^2} < 0$, 
$\displaystyle B_{nd}=-\frac{16 b \bar x ^5 (15+b \bar x +7 \bar x ^2)}{1+\bar x ^2} < 0$, 
and $C_{nd}<0$.
Therefore, $P_{nd}(\tau) < 0$ and 
$\bar {\mathbf{x}}^{(d)}$ becomes spiral in eq.(\ref{eqn:4-3}) 
for any $\tau>0$.
Next, we consider its stability.
From eq.(\ref{eqn:stable_ineq_ab}), 
the parameters $\alpha(\bar {\mathbf{x}})$ and $\beta(\bar {\mathbf{x}})$ can be calculated as
\begin{eqnarray}
    \alpha(\bar {\mathbf{x}}) & = & \frac{b \bar x ^2}{1+\bar x ^2}(3 \bar x ^2-5) = \frac{3}{25} \frac{b \bar x ^2}{1 + \bar x ^2} \left( a^2 - \frac{125}{3}\right), \nonumber \\
    \beta(\bar {\mathbf{x}}) & = & 3 \bar x ^3-b \bar x ^2 -5\bar x \nonumber\\
    								  & = & -\bar{x}^{2} \left\{ b - \frac{3}{5a} \left( a^2 - \frac{125}{3}\right) \right\}.
    \label{eqn:4-2-ab}
\end{eqnarray}
If $\displaystyle \alpha(\bar {\mathbf{x}}) < 0$, or $\displaystyle a^2 - \frac{125}{3} < 0$, 
then $\beta(\bar {\mathbf{x}})<0$ always holds 
from eq.(\ref{eqn:4-2-ab}) and $\bar {\mathbf{x}}^{(d)}$ is stable 
in eq.(\ref{eqn:4-3}) based on [St-2].
If $\displaystyle \alpha(\bar {\mathbf{x}}) > 0$, 
it is found from [St-1] that $\bar {\mathbf{x}}^{(d)}$ is stable
for $0<\tau< \gamma$ and unstable for $\tau > \gamma$, 
where $\gamma
=\displaystyle-\frac{(25+a^2)(3a^2-5ab-125)}{5ab(3a^2-125)}$.
When $3a^2-5ab-125 < 0$, $\gamma$ becomes positive.
Therefore, the Neimark-Sacker bifurcation surface 
is given as $\tau = \gamma$, namely   
%
%
%
%
%
\begin{equation}
    \tau = -\frac{(25+a^2)(3a^2-5ab-125)}{5ab(3a^2-125)}.
    \label{eqn:4-5}
\end{equation}
%
%
%
%
Figure \ref{Fig.LE_bif_ab_curve} shows the projection of the bifurcation surface, eq.(\ref{eqn:4-5}), on the $ab$-plane.
The Neimark-Sacker bifurcation surface exists in the region where  
$\displaystyle a > 5 \sqrt {\frac{5}{3}}$ and 
$\displaystyle b > \frac{3}{5}a-\frac{25}{a}$. 
%
%
%
When $\tau \to 0$, eq.(\ref{eqn:4-5}) gives 
$\displaystyle b=\frac{3}{5}a-\frac{25}{a}$, which coincides with 
the bifurcation line of eq.(\ref{eqn:4-1}).
The region for the unstable fixed point increases as $\tau$ increases, and
in the limit of $\tau \to \infty$, the $ab$-plane is divided into the stable and unstable regions by $\displaystyle a=5\sqrt{\frac{5}{3}}$.
%
%
%

Finally, we comment on the relevance of the tropical discretization to the non-standard finite difference scheme \cite{Mickens1994,Alexander2005}. 
%
%
The non-standard finite difference scheme is also known 
as a positivity-preserving difference scheme
and treats the set of differential equations 
\begin{eqnarray}
    \frac{d x}{dt}  =  f_1(x, y) - x h_1(x, y), 
    \frac{d y}{dt}  =  f_2(x, y) - y h_2(x, y),
    \label{eqn:ns1}
\end{eqnarray}
%
where $f_i,h_i$ are positive smooth functions ($i = 1, 2$).
Their discretized equations are given as
%
%
\begin{eqnarray}   
    x_{n+1} & = & 
       \frac{x_n + \tau f_1(x_n, y_n)}{1 + \tau h_1(x_n, y_n)}, \nonumber\\
    y_{n+1} & =  &
        \frac{x_n + \tau f_2(x_{n+1}, y_n)}{1 + \tau h_2(x_{n+1}, y_n)}.
    \label{eqn:ns2}
\end{eqnarray}
%
%
%
Comparing eq.(\ref{eqn:ns2}) with eq.(\ref{eqn:1-2}), 
we find $x$ in the functions $f_{2}(x, y)$ and $h_{2}(x, y)$ becomes $x_{n+1}$ instead of $x_{n}$.
Then the types and stability of the fixed points 
for eq.(\ref{eqn:ns2}) are consider to have different properties from the present ones.
The dynamical properties of eq.(\ref{eqn:ns2}) have been already discussed in the previous study\cite{Alexander2005}, but its relationship to the present results is not clear and will be the subject of future work.
%
%

\section{Conclusion}

Relationship of the dynamical properties in the vicinity of fixed points between two-dimensional continuous dynamical systems and their tropically discretized ones has been studied with the aid of the linear stability analysis.
The conditions under which the dynamical structures of the original continuous dynamical systems are retained are given by the signs of $\Delta$, $P_{sd}(\tau)$, and $P_{nd}(\tau)$.
Moreover, we identify stability of the fixed points in the discrete dynamical system as the conditions for $\tau$.
Our general results are successfully applicable to the Sel'kov model and the Lengyel-Epstein model.
Especially in the case of the Sel'kov model, our present analytical results are consistent with our previous numerical ones.
%
%


\noindent
{\bf Acknowledgments}

The authors are grateful to 
Prof. D. Takahashi, 
Prof. T. Yamamoto, and Prof. Emeritus A. Kitada 
at Waseda University, 
Associate Prof. K. Matsuya at Musashino University, Prof. M. Murata at Tokyo University of Agriculture and Technology 
for useful comments and encouragements. 
This work was supported by JSPS
KAKENHI Grant Numbers 22K13963 and 22K03442.

\end{document}